\documentclass{ws-ijmpa}

\begin{document}

\markboth{Laura Elisa Marcucci}
{Muon capture on deuteron and $^3$He: a personal review}

\catchline{}{}{}{}{}
%
%%%%%%%%%%%%%%%%%%%%%%%%%%%%%%%%%%%%%%%%%%%%%%%%%%%%%%%%%%%%%%%%%%%%

\title{MUON CAPTURE ON DEUTERON AND $^3$HE: A PERSONAL REVIEW}

\author{LAURA ELISA MARCUCCI}

\address{Department of Physics ``E. Fermi'', University of Pisa, 
and INFN-Pisa, \\
Largo Bruno Pontecorvo, 3, 
Pisa, I-56127,
Italy\\
laura.marcucci@df.unipi.it}

\maketitle

\begin{history}
\received{Day Month Year}
\revised{Day Month Year}
\end{history}

\begin{abstract}
The present status of theoretical and experimental studies of
muon capture reactions on light nuclei is reviewed. In particular, 
the recent results for the two reactions 
$^2$H($\mu^-,\nu_\mu$)$nn$ and $^3$He($\mu^-,\nu_\mu$)$^3$H are
presented, and the 
unresolved discrepancies among different measurements and
calculations, open problems,
and future developments are discussed.

\end{abstract}

\ccode{PACS numbers: 23.40.-s,21.45.-v,27.10.+h}

\section{Introduction}
\label{sec:intro}

When negative muons pass through matter, they can be captured into high-lying
atomic orbitals.  Then, in a time-scale of the order of 10$^{-13}$ s, they 
cascade down into the 1$s$ orbit, through Auger processes with
atomic electrons and the emission of X-rays. 
At this point,
two competing processes occur: one is ordinary decay
\begin{equation}
\mu^-\rightarrow e^-\,+\,\overline{\nu}_e\,+\, \nu_\mu \ , 
\label{eq:ordecay}
\end{equation}
and the other is the (weak) capture by the nucleus 
\begin{equation}
\mu^-\, + \, A(Z,N)\rightarrow \nu_\mu \,+\, A(Z-1,N+1) \ ,
\label{eq:capture}
\end{equation}
which can take place from any of the two initial 
hyperfine states, $f=J_i\pm 1/2$ ($J_i$ is the spin of the 
initial nucleus $A(Z,N)$).
Apart from tiny corrections due to bound-state effects~\cite{Cza00}, 
the decay rate is essentially the same as for a free muon.
In light nuclei, this 
is much larger than the rate for capture, which 
proceeds predominantly through the basic process 
\begin{equation}
\mu^-\,+\, p \rightarrow n\,+\, \nu_\mu \ ,
\label{eq:basicproc}
\end{equation}
induced by the exchange of a $W^+$ boson.
Its rate is expected to be proportional to the number of protons in
the nucleus and to the probability of finding the muon at the nucleus.
Since the semi-leptonic weak nuclear interaction is effectively a 
contact interaction, this probability scales like the
square of the atomic 1$s$ wave function evaluated at 
the origin~\cite{Pri59}, proportional to $Z^{\,3}$.
The capture rate, therefore, scales roughly 
like $Z^{\, 4}$. 
It is only for nuclei with $Z \geq 12$ that the nuclear capture rate
becomes comparable with
the decay rate. Muonic capture on light nuclei are therefore
experimentally challenging processes. However, they are
preferred under the theoretical point of view, 
as the nuclear effects can be easier and more accurately taken into account,
and informations on the basic process of Eq.~(\ref{eq:basicproc}) can be better
extracted. For instance, 
muon capture on hydrogen and hydrogen isotopes
is, in principle, best suited to obtain
informations on the matrix element of the (charge-changing) single-nucleon
weak current
\begin{equation}
j^\mu=
\overline{u_p} \left[F_1(q^2)\gamma^\mu + F_2(q^2) \frac{i\sigma^{\mu\nu}q_\nu}{2M_N} 
-G_A(q^2) \gamma^\mu\gamma^5-G_{PS}(q^2)\frac{q^\mu\gamma^5}{2M_N}\right]u_n \ ,
\label{eq:wknuclcurr}
\end{equation}
but is experimentally the hardest process. 
Note that in Eq.~(\ref{eq:wknuclcurr}) we have 
ignored 
contributions from second-class currents~\cite{Wei58}, for which
there is presently no firm experimental evidence~\cite{Sev06},
and we have indicated with $u_p$ ($u_n$) the proton
(neutron) spinor, with $M_N$ the nucleon mass and with
$q^2$ the four-momentum transfer; $\gamma^\mu$ and $\gamma^5$
are the standard Dirac matrices, $\sigma^{\mu\nu}=i/2 [\gamma^\mu,\gamma^\nu]$ 
and $\overline{u_p}=u_p^\dagger \gamma_0$~\cite{Bjo64}.
Of the four form factors of Eq.~(\ref{eq:wknuclcurr}), 
$F_1(q^2)$ and $F_2(q^2)$ are related to the
isovector electromagnetic form factors of the nucleon by the
conserved-vector-current (CVC) constraint. They
are well known over a wide range of momentum transfers $q^2$
from elastic electron scattering data on the nucleon~\cite{FF}.  
The axial form factor $G_A(q^2)$
is also quite well known: its value at vanishing $q^2$, 
$g_A=1.2695\pm 0.0029$, is
from neutron $\beta$-decay~\cite{PDG}, while its $q^2$-dependence
is parametrized as 
\begin{equation}
G_A(q^2)=g_A/( 1-q^2/\Lambda_A^2)^2 \ , 
\label{eq:gaq2}
\end{equation}
with $\Lambda_A= 1$ GeV from an analysis 
of pion electro-production data~\cite{Ama79}
and direct measurements of the reaction 
$p\,+\,\nu_\mu\rightarrow n\,+\, \mu^+$~\cite{Kit83}.
Note that a considerably larger value
$\Lambda_A=1.35$ GeV is obtained from current analyses of neutrino
quasi-elastic scattering data on nuclear targets~\cite{Juszczak10}.  However,
these analyses are based on rather crude models of nuclear structure 
(Fermi gas or a local density approximation 
of the nuclear matter spectral function)
and on simplistic treatments of the reaction mechanism.
Also, some discrepancies exist on the neutron $\beta$-decay lifetime,
as the world average value used here~\cite{PDG} differs by 6.5
standard deviations from the results obtained from 
gravitationally trapped ultra-cold neutrons~\cite{Ser08-10}. A discussion
of this point is however well beyond the subject of the 
present review.

The induced pseudoscalar form factor $G_{PS}(q^2)$ is the least known
of the four form factors of Eq.~(\ref{eq:wknuclcurr}).  
The MuCap collaboration at Paul Scherrer Institute (PSI) 
has recently reported a precise measurement 
of the rate for reaction~(\ref{eq:basicproc}) 
in the singlet hyperfine state ($f=0$): 
$725.0 \pm 13.7 ({\rm stat}) \pm 10.7 ({\rm syst})$ sec$^{-1}$~\cite{MuCap}.
Based on this value, 
an indirect \lq\lq experimental\rq\rq determination of $G_{PS}$ at the
momentum transfer $q_0^2=-0.88\, m_\mu^2$ relevant for muon 
capture on hydrogen has been given~\cite{Cza07}, 
$G^{\,{\rm EXP}}_{PS}(q_0^2)=7.3\pm 1.2$, by using
for the remaining form factors the values discussed above and by evaluating
electroweak radiative corrections.  These are found to be sizable,
of the order of $\sim 3$ \%.
Theoretical predictions for the induced pseudoscalar form factor were derived long ago
based on the notion of a partially conserved axial current (PCAC) and
pion dominance, and were later refined by evaluating leading-order corrections
to the PCAC result with current algebra techniques~\cite{Adl66}.
More recently, these predictions have been re-derived in chiral perturbation
theory ($\chi$PT)~\cite{GPth1,GPth2}, finding 
\begin{equation}
G_{PS}^{\rm TH}(q^2)=\frac{2 m_\mu g_{\pi p n}f_\pi}{m_\pi^2-q^2} -\frac{1}{3}
g_A m_\mu M_N r_A^2 \ ,
\label{eq:gpsth}
\end{equation}
where $g_{\pi p n}=13.05\pm 0.20$ is the $\pi NN$ coupling constant, 
$f_\pi=92.4\pm 0.4$ MeV is the pion decay constant, 
and $r_A=0.43\pm 0.03$ fm$^{2}$ is the 
axial radius of the nucleus, related to $\Lambda_A$ of
Eq.~(\ref{eq:gaq2}) as $\Lambda_A^2=12/r_A^2$.
For $q_0^2=-0.88\, m_\mu^2$, 
$G^{\rm TH}_{PS}(q_0^2)=8.2 \pm 0.2$~\cite{GPth1}. 
To be noticed that the evaluation of electroweak radiative 
corrections~\cite{Cza07} for muon capture on hydrogen
is crucial for bringing $G^{\,{\rm EXP}}_{PS}$ 
within less than 1$\sigma$ of $G^{\rm TH}_{PS}$.

Besides their relevance for extracting 
informations on single-nucleon weak current
form factors, muon captures on light nuclei also
provide a testing ground for 
the theoretical frameworks used to study those
reactions of astrophysical interest
whose rates cannot be measured experimentally,
and for which one has to rely exclusively on theory~\cite{SFII}.
In fact, the same nuclear wave functions and,
indirectly, the same model for the nuclear 
interactions from which these are obtained, and
the same nuclear weak current can be used to study
neutrino reactions in light nuclei~\cite{Gazetal},  
weak proton captures on proton 
and $^3$He (the so-called $pp$ and $hep$ reactions)~\cite{Sch98,Mar00,Par03},
and muon captures on light nuclei. 

In the present review,
we focus our attention on the following captures:
\begin{eqnarray}
&&\mu^-+d\rightarrow n+n+\nu_\mu \ , \label{eq:mud}\\
&&\mu^-+\,^3{\rm He}\rightarrow\,^3{\rm H}+\nu_\mu \ . \label{eq:mu3}
\end{eqnarray}
Muon capture on $^3$He can also occur through the two- ($nd$) and
three-body ($nnp$) breakup channels of $^3{\rm H}$. However, 
the branching ratios of these two processes are 
20 \% and 10 \%, respectively, and experimental and theoretical
work on them is quite limited. They will not be discussed here.
A comprehensive and detailed description of
these reactions, and more in general of the physics of muon capture
and the problem of the induced pseudoscalar form factor, can be 
found in Refs.~\refcite{Mea01},~\refcite{Gor04} and~\refcite{Kam10}.

The observables of interest for muon capture reactions
are the capture rates. 
In reaction~(\ref{eq:mud}), 
the stopped muons can in principle be captured from the two hyperfine states, 
$f=1/2$ or $3/2$. However, it is known that
capture takes place practically uniquely from the doublet hyperfine 
state~\cite{Mea01,Gor04}. We will therefore consider only 
the doublet capture rate $\Gamma^D$.
In reaction~(\ref{eq:mu3}), instead, 
a difference in the capture rates
between the hyperfine states is not expected: 
a hyperfine transition is highly unlikely, due to the 
energy difference between the hyperfine states. We will therefore
consider the total capture rate $\Gamma_0$. In the next section
we will briefly discuss the experimental situation, while
the formalism to derive these observables and
the most recent theoretical calculations will be
presented in Sec.~\ref{sec:theory}.
Some concluding remarks are given in Sec.~\ref{sec:concl}.

\section{Experimental situation}
\label{sec:exp}

The first attempt to measure $\Gamma^D$ was carried out over
forty years
ago by Wang {\it et al.}~\cite{Wan65}. Using a liquid 
mixed H$_2$/D$_2$ target, they obtained 
$\Gamma^D=365 \pm 96$ s$^{-1}$.
A few years later, Bertin {\it et al.} 
measured $\Gamma^D=445 \pm 60$ s$^{-1}$~\cite{Ber73},
using a gas mixed H$_2$/D$_2$
target, and assuming a pure doublet mix of $\mu d$ spin states.
However,
a subsequent study of hyperfine depopulation in a H$_2$/D$_2$ 
mixture~\cite{Bre81}
has failed to support this assumption, and therefore
the Bertin {\it et al.} result is considered controversial~\cite{Gor04}.
The most recent measurements have been
performed in the eighties by Bardin {\it et al.}~\cite{Bar86}, and
Cargnelli {\it et al.}~\cite{Car86}. They both used pure deuterium,
so that the $\mu d + d$ collision rate is sufficient to
fully depopulate the $f=3/2$ hyperfine state.  
Furthermore, Bardin {\it et al.} used a liquid target and the
lifetime method, i.e. compared the negative and positive muon lifetime
when stopped in deuterium. For positive muons, the lifetime is the inverse
of the muon decay rate, while for negative muons the lifetime is the
inverse of the sum of the muon decay rate and the muon capture rate.
The lifetime difference thus determines the capture rate,
assuming, according to the CPT theorem, 
the positive and negative muon decay rate identical.
Cargnelli {\it et al.}, instead, used a gas target and the 
neutron method, i.e. directly detected the recoil neutrons, which is 
obviously quite challenging. These two measurements 
gave $\Gamma^D=470 \pm 29$ s$^{-1}$~\cite{Bar86}
and $\Gamma^D=409 \pm 40$ s$^{-1}$~\cite{Car86}.
In conclusion, all the measurements available until now, 
while consistent with each other, are not 
very precise, since the errors are in the 6$\div$10 \% range. 
However, there
is hope to have this situation clarified by the MuSun 
Collaboration~\cite{And10,Kam10}, with their on-going 
experiment at PSI, which should reach a precision of 1.5 \%.
The gain in experimental precision relies on the fundamental 
techniques developed for the MuCap experiment~\cite{MuCap}. Muons
will be stopped in an active gas target consisting of a 
cryogenic ionization chamber operated as time
projection chamber with ultra-pure deuterium. The muon stopping point
will be reconstructed in 3 dimensions, and this will 
eliminate the otherwise overwhelming
background from muon stops in wall materials. 
The capture rate will then be determined using the lifetime
technique.

The experimental situation for muon capture on $^3$He is much clearer.
After a first set of measurements in the early 
sixties by Falomkin {\it et al.}~\cite{Fal63},
Zaimidoroga {\it et al.}~\cite{Zai63}, 
Auerbach {\it et al.}~\cite{Aue65} and
Clay {\it et al.}~\cite{Cla65}, 
a very precise determination was performed 
by Ackerbauer {\it et al.}~\cite{Ack98} in the late nineties. 
The basic method involves counting the numbers of muon stops and
$^3$He recoils, when a beam of muons is stopped in $^3$He. 
Ackerbauer {\it et al.}
used a gas ionization chamber, which allowed a better separation
of muon and $^3$He signals.
The measured total capture rate $\Gamma_0$, corresponding
to a statistical average population of the 4 different hyperfine states, 
is $1496 \pm 4$ s$^{-1}$~\cite{Ack98},
a value consistent with those of the earlier measurements, 
but with a factor 10 of improvement in the experimental accuracy.

If the hyperfine structure of the
($\mu,^3$He) system is taken into account
and the direction of the recoiling triton is
detected, there are, in addition to the
total capture rate, other observables, i.e. angular
correlation parameters or so-called recoil asymmetries, 
which are more sensitive than the capture rate itself 
to the value of the induced pseudoscalar form factor $G_{PS}(q^2)$.
A first attempt to measure the recoil asymmetry 
has been made by Souder {\it et al.} at TRIUMF~\cite{Sou98}.
They used a $^3$He ionization chamber to stop the
incoming muons, re-polarize the ($\mu,^3$He) system and track the triton
recoils.
They obtained for the vector asymmetry $A_v$
the value of 0.63 $\pm$ 0.09 (stat.)$^{+0.11}_{-0.14}$ (syst.).
This experimental result, which to our knowledge represents
the first measurement of this
observable, is affected by large
systematic uncertainties.  Therefore, a
comparison between theory and experiment
would not be particularly meaningful. Thus,
further experimental work is highly recommended.

\section{Theoretical calculations}
\label{sec:theory}

Before discussing the results of the different theoretical
calculations for the capture rates of reactions~(\ref{eq:mud})
and~({\ref{eq:mu3}), we present in the following subsection
the formalism used in the calculation of the observables 
under consideration.

\subsection{Theoretical formalism}
\label{subsec:form}

The muon capture on deuteron and $^3$He is induced by the weak interaction
Hamiltonian~\cite{Wal95}
\begin{equation}
H_{W}={\frac{G_{V}}{\sqrt{2}}} \int {\rm d}{\bf x}
\, l_{\sigma}({\bf x}) j^{\sigma}({\bf x}) \ , \label{eq:hw}
\end{equation}
where $G_{V}$ is the Fermi coupling constant,
$G_{V}$=1.14939 $\times 10^{-5}$ GeV$^{-2}$ as
obtained from an analysis of $0^+ \rightarrow 0^+$ $\beta$-decays~\cite{Har90},
and $l_\sigma$ and $j^\sigma$ are the leptonic and
hadronic current densities, respectively.  The former
is given by
\begin{equation}
l_{\sigma}({\bf x}) =\, 
{\rm e}^{-{\rm i} {\bf k}_\nu \cdot {\bf x} } \,
{\overline{u}}({\bf k}_\nu,h_\nu)\,\gamma_{\sigma}\, (1-\gamma_5)
\psi_{\mu}({\bf x},s_{\mu}) \>\>\>,
\label{eq:lepc}
\end{equation}
where $\psi_\mu({\bf x},s_\mu)$ is the ground-state
wave function of the muon in the Coulomb field of the nucleus in the
initial state, and $u({\bf k}_\nu,h_\nu)$ is the spinor of
a muon neutrino with momentum ${\bf k}_\nu$, energy
$E_\nu$ (=$k_\nu$), and helicity $h_\nu$.  While in principle
the relativistic solution of the Dirac equation could be
used, in practice it suffices to approximate
\begin{eqnarray}
\psi_\mu({\bf x},s_\mu) &\simeq& 
\psi_{1s}(x) \chi(s_\mu) \equiv 
\psi_{1s}(x) u({\bf k}_\mu,s_\mu) \nonumber \\
{\bf k}_\mu &\rightarrow& 0 \>\>\>,
\label{eq:psimu}
\end{eqnarray}
since the muon velocity
$v_\mu \simeq Z \alpha \ll 1$ ($\alpha$ is
the fine-structure constant and $Z$=1 or 2 for deuteron or
$^3$He, respectively).  Here
$\psi_{1s}(x)$ is the $1s$ solution of the Schr\"odinger
equation and, since the muon is essentially at rest,
it is justified to replace the two-component
spin state $\chi(s_\mu)$ with the four-component
spinor $u({\bf k}_\mu,s_\mu)$
in the limit ${\bf k}_\mu \rightarrow 0$.  This will
allow us to use standard techniques to carry out
the spin sum over $s_\mu$ at a later stage. 

In order to account for the hyperfine structure
in the initial system, the muon and deuteron or $^3$He
spins are coupled to states with total spin $f$, equal to 1/2 or 3/2 in the
deuteron case, and to 0 or 1 in the $^3$He case.  
The transition amplitude can then be conveniently
written as~\cite{Mar11a}
\begin{eqnarray}
T_W (f,f_z;s_1,s_2,h_\nu) &\equiv&
\langle nn, s_1, s_2; \nu, h_\nu \,|\, H_W \,|\,
(\mu,d);f,f_z \rangle
\nonumber \\
&\simeq& {\frac{G_V} {\sqrt{2}}} \psi_{1s}^{\rm av}
\sum_{s_\mu s_d}
\langle {\frac{1}{2}}s_{\mu}, 1 s_d | f f_z \rangle\,
l_\sigma(h_\nu,\,s_\mu)\, \nonumber \\
&&
\langle \Psi_{{\bf p}, s_1 s_2}(nn) | j^{\sigma}({\bf q}) |
\Psi_d(s_d)\rangle \ , \label{eq:h2ffz}
\end{eqnarray}
for the muon capture on deuteron, where ${\bf p}$ is the $nn$ relative
momentum, and~\cite{Mar02}
\begin{eqnarray}
T_W (f,f_z;s^\prime_{3},h_\nu) &\equiv&
\langle ^3{\rm H}, s^\prime_{3}; \nu, h_\nu \,|\, H_W \,|\,
(\mu,^3\!{\rm He});f,f_z \rangle
\nonumber \\
&\simeq& {\frac{G_V}{\sqrt{2}}} \psi_{1s}^{\rm av}
\sum_{s_\mu s_3}
\langle {\frac{1}{2}}s_{\mu}, {\frac{1}{2}} s_3 | f f_z \rangle\,
l_\sigma(h_\nu,\,s_\mu)\, \nonumber\\
&&
\langle \Psi_{^3{\rm H}}(s^\prime_{3}) | j^{\sigma}({\bf q}) |
\Psi_{^3{\rm He}} (s_3)\rangle \ , \label{eq:h3ffz}
\end{eqnarray}
for muon capture on $^3$He.
In Eqs.~(\ref{eq:h2ffz}) and~(\ref{eq:h3ffz}) we have defined 
\begin{equation}
l_\sigma(h_\nu,\,s_\mu) \equiv
{\overline{u}}({\bf k}_\nu,h_\nu)\,\gamma_{\sigma}\, (1-\gamma_5)
u({\bf k}_\mu,s_{\mu}) \>\>\>,
\label{eq:lsigma}
\end{equation}
and the Fourier transform of the nuclear weak current
has been introduced as
\begin{equation}
j^\sigma({\bf q})=\int {\rm d}{\bf x}\,
{\rm e}^{ {\rm i}{\bf q} \cdot {\bf x} }\,j^\sigma({\bf x})
\equiv (\rho({\bf q}),{\bf j}({\bf q}))
\label{eq:jvq} \>\>\>,
\end{equation}
with the leptonic momentum transfer ${\bf q}$ defined
as ${\bf q} = {\bf k}_\mu-{\bf k}_\nu \simeq -{\bf k}_\nu$.
The function $\psi_{1s}(x)$ has been
factored out from the matrix element of $j^{\sigma}({\bf q})$
between the initial and final states. For muon capture on deuteron,
$\psi_{1s}^{\rm av}$ is approximated as~\cite{Wal95}
\begin{equation}
|\psi_{1s}^{\rm av}|^2 \equiv\,  |\psi_{1s}(0)|^2\,=\,
{\frac{(\alpha\, \mu_{\mu d})^3}{\pi}} \ ,
\label{eq:psimud}
\end{equation}
where $\psi_{1s}(0)$ denotes the Bohr wave function
for a point charge $e$ evaluated at the origin, and
$\mu_{\mu d}$ is the reduced mass of the $(\mu,d)$ system.
For muon capture on $^3$He,
$\psi_{1s}^{\rm av}$ is approximated as~\cite{Mar02}
\begin{equation}
|\psi_{1s}^{\rm av}|^2 \equiv\,  
{\cal {R}}\,{\frac{(2\,\alpha\, \mu_{\mu ^3{\rm He}})^3}{\pi}} \ ,
\label{eq:psimu3}
\end{equation}
where in this case 
$\mu_{\mu ^3{\rm He}}$ is the reduced mass of the ($\mu,^3$He) system,
and the factor ${\cal {R}}$ approximately accounts
for the finite extent of the nuclear charge
distribution~\cite{Wal95}. 
This factor is defined as
\begin{equation}
{\cal R}=\frac{|\psi_{1s}^{\rm av}|^2}{|\psi_{1s}(0)|^2}\ ,
\label{eq:rf}
\end{equation}
with
\begin{equation}
\psi_{1s}^{\rm av}=\frac
{\int d{\bf x} \,{\rm e}^{{\rm i}{\bf q}\cdot{\bf x}}\psi_{1s}(x)\rho(x)}
{\int d{\bf x} \,{\rm e}^{{\rm i}{\bf q}\cdot{\bf x}}\rho(x)} \ ,
\label{eq:psiav}
\end{equation}
where $\rho(x)$ is the $^3$He charge density. 
It has been calculated
explicitly in Ref.~\refcite{Mar11a} by using the charge densities 
corresponding to two realistic Hamiltonian models, 
the AV18/UIX and N3LO/N2LO (see below), 
and has been found for both models to be within a percent
of 0.98, the value obtained from the experimental charge
density and commonly adopted in the literature~\cite{Wal95}.

In the case of muon capture on deuteron, the final state wave
function is expanded in partial waves as
\begin{equation}
\Psi_{{\bf p},s_1,s_2}(nn)=4\pi\sum_{S} \langle \frac{1}{2} s_1,\frac{1}{2}s_2 |
S S_z \rangle 
\sum_{L J J_z}{\rm i}^L Y^*_{LL_z}({\hat{\bf p}}) 
\langle S S_z, L L_Z | J J_z\rangle \,\overline{\Psi}_{nn}^{LSJJ_z}(p) \>\> ,
\label{eq:psinnpw}
\end{equation}
where $\overline{\Psi}_{nn}^{LSJJ_z}(p)$ is the $nn$ wave function.
The calculation is typically restricted to $J\leq 2$ and
$L\leq 3$, since it has been proven that higher order partial waves
give negligible contributions~\cite{Mar11a}.
Therefore, in spectroscopic notation, 
only the $^1S_0$, $^3P_0$, $^3P_1$, $^3P_2$--$^3F_2$, 
and $^1D_2$ partial waves are considered.

Now, standard techniques~\cite{Mar00,Wal95} are
used to carry out the multipole expansion
of the weak charge, $\rho({\bf q})$, and current,
${\bf j}({\bf q})$, operators. For muon capture on 
deuteron, we find
\begin{eqnarray}
\langle \overline{\Psi}_{nn}^{LSJJ_z}(p) | \rho({\bf q}) | \Psi_d(s_d) \rangle 
&=&
\sqrt{4\pi}\sum_{\Lambda \geq 0}\sqrt{2\Lambda+1}\,\,{\rm i}^\Lambda
\frac{\langle 1 s_d, \Lambda 0 | J J_z\rangle}{\sqrt{2J+1}} C_\Lambda^{LSJ}(q) \ ,
\label{eq:c2} \\
\langle \overline{\Psi}_{nn}^{LSJJ_z}(p) | j_z({\bf q}) | \Psi_d(s_d) \rangle 
&=&
-\sqrt{4\pi}\sum_{\Lambda \geq 0}\sqrt{2\Lambda+1}\,\,{\rm i}^\Lambda
\frac{\langle 1 s_d, \Lambda 0 | J J_z\rangle}{\sqrt{2J+1}} L_\Lambda^{LSJ}(q) \ ,
\label{eq:l2} \\
\langle \overline{\Psi}_{nn}^{LSJJ_z}(p) | j_\lambda({\bf q}) | 
\Psi_d(s_d) \rangle 
&=&
\sqrt{2\pi}\sum_{\Lambda \geq 1}\sqrt{2\Lambda+1}\,\,{\rm i}^\Lambda
\frac{\langle 1 s_d, \Lambda -\lambda | J J_z\rangle}{\sqrt{2J+1}} 
\nonumber \\
&&
[-\lambda M_\Lambda^{LSJ}(q)+E_\Lambda^{LSJ}(q)]\ ,
\label{eq:em2}
\end{eqnarray}
where $\lambda=\pm 1$, and $C_{\Lambda}^{LSJ}(q)$, $L_{\Lambda}^{LSJ}(q)$,
$E_{\Lambda}^{LSJ}(q)$ and $M_{\Lambda}^{LSJ}(q)$ denote the reduced matrix 
elements (RME's) of the Coulomb ($C$), longitudinal ($L$), transverse
electric ($E$) and transverse magnetic ($M$) multipole operators, as defined
in Ref.~\refcite{Mar00}. Since the weak charge/current
operators have scalar/polar-vector $(V)$ and 
pseudo-scalar/axial-vector $(A)$
components, each multipole consists of the sum of $V$ and $A$ terms,
having opposite parity under space inversion~\cite{Mar00}.
The contributing multipoles for the $S$-, $P$-, and $D$-channels
mentioned above in muon capture on deuteron are given in Table~\ref{tab:rme2},
where the superscripts $LSJ$ have been dropped.
\begin{table}[b]
\tbl{Contributing
multipoles in muon capture on deuteron, for all the $nn$ partial waves
with $J\leq 2$ and $L\leq 3$.
The spectroscopic notation is used. See text for further explanations.}
{\begin{tabular}{@{}cc@{}} \toprule
Partial wave & Contributing multipoles \\ \colrule
$^1S_0$ & $C_1(A)$, $L_1(A)$, $E_1(A)$, $M_1(V)$ \\
$^3P_0$ & $C_1(V)$, $L_1(V)$, $E_1(V)$, $M_1(A)$ \\
$^3P_1$ & $C_0(A)$, $L_0(A)$, \\
        & $C_1(V)$,  $L_1(V)$,  $E_1(V)$,  $M_1(A)$,\\
        & $C_2(A)$,  $L_2(A)$,  $E_2(A)$,  $M_2(V)$ \\
$^3P_2$--$^3F_2$
        & $C_1(V)$,  $L_1(V)$,  $E_1(V)$,  $M_1(A)$,\\
        & $C_2(A)$,  $L_2(A)$,  $E_2(A)$,  $M_2(V)$,\\
        & $C_3(V)$,  $L_3(V)$,  $E_3(V)$,  $M_3(A)$ \\
$^1D_2$ 
        & $C_1(A)$,  $L_1(A)$,  $E_1(A)$,  $M_1(V)$,\\
        & $C_2(V)$,  $L_2(V)$,  $E_2(V)$,  $M_2(A)$,\\
        & $C_3(A)$,  $L_3(A)$,  $E_3(A)$,  $M_3(V)$ \\ 
\botrule
\end{tabular} \label{tab:rme2}}
\end{table}
In the case of muon capture on $^3$He, explicit expressions for the
multipole operators are given by~\cite{Mar02}
\begin{equation}
\langle \Psi_{^3{\rm H}}(s^\prime_{3})  | \rho({\bf q}) |
\Psi_{^3{\rm He}}(s_{3})  \rangle = \sqrt{2\pi}
\sum_{l=0,1}\sqrt{2l + 1}\,{\rm i}^l \,
d_{m,0}^l(-\theta)\,
\langle {\frac{1}{2}}s_{3}, l\,m | {\frac{1}{2}}s^\prime_{3}\rangle
\, C_{l}(q) \ , \label{eq:c}
\end{equation}
\begin{equation}
\langle \Psi_{^3{\rm H}}(s^\prime_{3}) | j_z({\bf q}) |
\Psi_{^3{\rm He}}(s_{3})  \rangle =-\sqrt{2\pi}
\sum_{l=0,1}\sqrt{2l + 1}\,{\rm i}^l \,
d_{m,0}^l(-\theta)\,
\langle {\frac{1}{2}}s_{3}, l\,m | {\frac{1}{2}}s^\prime_{3}\rangle
\, L_{l}(q) \ , \label{eq:l}
\end{equation}
\begin{equation}
\langle \Psi_{^3{\rm H}}(s^\prime_{3}) | j_\lambda({\bf q}) |
\Psi_{^3{\rm He}}(s_{3})  \rangle = \sqrt{3 \pi}
\, \,{\rm i}  \, d_{m,-\lambda}^1(-\theta)\,
\langle {\frac{1}{2}}s_{3}, 1 \,m | {\frac{1}{2}}s^\prime_{3}\rangle
[-\lambda M_1(q)+E_1(q)] \ , \label{eq:em}
\end{equation}
where $m$=$s_3^\prime-s_3$, and 
the $d^l_{m,m^\prime}$
are rotation matrices in the standard notation of Ref.~\refcite{Edm57}.
Applying parity
and angular momentum selection rules, it has been shown~\cite{Mar02}
that the only contributing RME's are 
$C_0(V)$, $C_1(A)$, $L_0(V)$, $L_1(A)$, $E_1(A)$, and $M_1(V)$.

The total capture rate for the two reactions under consideration
is then defined as
\begin{equation}
d\Gamma = 2\pi\delta(\Delta E) \overline{|T_W|^2} \times({\rm phase \, space})
\ ,
\label{eq:dgamma}
\end{equation}
where $\delta(\Delta E)$ is the energy-conserving $\delta$-function, 
and the phase space is 
$d{\bf p}\,d{\bf k}_\nu/(2\pi)^6$ 
for reaction~(\ref{eq:mud}) and just
$d{\bf k}_\nu/(2\pi)^3$ for reaction~(\ref{eq:mu3}).
The following notation has been introduced: (i) for muon capture
on deuteron
\begin{equation}
\overline{|T_W|^2} = \frac{1}{2f+1}\sum_{s_1 s_2 h_\nu}\sum_{f_z}
|T_W(f,f_z;s_1,s_2, h_\nu)|^2 \ ,
\label{eq:hw2}
\end{equation}
and the initial hyperfine state has been fixed to be $f=1/2$; 
(ii) for muon capture on $^3$He
\begin{equation}
\overline{|T_W|^2} = \sum_{s_3^\prime  h_\nu}\sum_{f  f_z}
P(f,f_z) |T_W(f,f_z;s^\prime_3, h_\nu)|^2 \ ,
\label{eq:hw3}
\end{equation}
where $P(f,f_z)$ is the probability of
finding the ($\mu,^3$He) system
in the total-spin state $f,f_z$ and
$P(f,f_z)= 1/4$ when the same probability
to the different hyperfine states is assigned. 

After carrying out the spin sums, 
the total rate and recoil asymmetry for muon capture on $^3$He 
are~\cite{Mar02}
\begin{eqnarray}
\Gamma_0 &=& G_V^2\, E_\nu^2\, \left(1-{\frac{E_\nu}{m_{^3{\rm H}}}}\right)
\,|\psi_{1s}^{\rm av}|^2 \nonumber \\
&&\Big[\,|C_0(V)-L_0(V)|^2\,+\,|C_1(A)-L_1(A)|^2 
+|M_1(V)-E_1(A)|^2 \,\Big] \ ,
\label{eq:g0}
\end{eqnarray}
with $E_\nu$ given by
\begin{equation}
E_\nu=\frac{(m_\mu + m_{^3{\rm He}})^2 -m_{^3{\rm H}}^2}{2 (m_\mu + m_{^3{\rm He}})}
 \ ,
\label{eq:enu3}
\end{equation}
and
\begin{equation}
A_v=1+\frac{
 2\, {\rm Im}\Big[ \big( C_0(V)-L_0(V)\big)
                         \big( C_1(A)-L_1(A)\big)^{*} \Big]
-|M_1(V)-E_1(A)|^2}
{|C_0(V)-L_0(V)|^2\,+\,|C_1(A)-L_1(A)|^2 
+|M_1(V)-E_1(A)|^2 }
\ . \label{eq:av} 
\end{equation}

In the case of muon capture on deuteron, the differential rate reads
\begin{equation}
\frac{d\Gamma_D}{dp}=E_\nu^2\, \left[ 1-{\frac{E_\nu}{(m_\mu + m_d)}}\right]
\,|\psi_{1s}^{\rm av}|^2 \frac{p^2d{\hat{\bf p}}}{8\pi^4}\,
\overline{|T_W|^2} \ ,
\label{eq:dgd}
\end{equation}
where
\begin{equation}
E_\nu=\frac{(m_\mu +m_d)^2 -4m_n^2-4 p^2}{2 (m_\mu +m_d)} \ .
\label{eq:enu2}
\end{equation}
In Eqs.~(\ref{eq:g0})--(\ref{eq:enu2}), $m_\mu$, $m_n$, $m_d$, $m_{^3{\rm H}}$,
$m_{^3{\rm He}}$ are the muon, neutron, deuteron, $^3$H and $^3$He masses.
The integration over ${\hat{\bf p}}$ in Eq.~(\ref{eq:dgd})
is performed numerically using Gauss-Legendre points.
A limited number of them, of the order of 10, 
is necessary to achieve convergence to better than 1 part in 10$^3$. In 
order to
calculate the total capture rate $\Gamma^D$, the differential 
capture rate
is plotted versus $p$, and numerically
integrated. Usually, about 30 points in $p$ are enough for this integration
in each partial wave~\cite{Mar11a}.

\subsection{Results}
\label{subsec:theory}

Theoretical work on reactions~(\ref{eq:mud}) and~(\ref{eq:mu3})
is just as extensive as the experimental one (see Sec.~\ref{sec:exp}).
A list of publications, updated to the late nineties, is given 
in Table 4.1 of Ref.~\refcite{Mea01},
in Ref.~\refcite{Gor04} and Ref.~\refcite{Kam10}.
Here, we limit our considerations to the calculations 
performed since the year 2000.
The starting point will be our studies of 
Refs.~\refcite{Mar11a} and~\refcite{Mar11b},
for both muon capture reactions under consideration, as, to our 
knowledge, are the most recent ones published.
These results will be compared 
with the calculations of 
Ando {\it et al.}~\cite{And02} and Ricci {\it et al.}~\cite{Ric10}
for reaction~(\ref{eq:mud}), and 
Gazit~\cite{Gaz08}
for reaction~(\ref{eq:mu3}). We will comment also on our early
study of reaction~(\ref{eq:mu3})~\cite{Mar02}, and on the results 
of Ho {\it et al.}\cite{Ho02} and Chen {\it et al.}\cite{Che05}. 
The most recent studies (available only as preprint)
of Refs.~\refcite{Mar11c} and~\refcite{Ada11}
will be also briefly discussed.

The theoretical results of Refs.~\refcite{Mar11a},~\refcite{And02}
and~\refcite{Ric10}
for the capture rate $\Gamma^D$ of 
reaction~(\ref{eq:mud}) from the initial doublet hyperfine state
and of Refs.~\refcite{Mar11a},~\refcite{Mar02}, and~\refcite{Gaz08}
for the total capture rate $\Gamma_0$ of
reaction~(\ref{eq:mu3}) 
are summarized in Tables~\ref{tab:res2} and~\ref{tab:res3},
respectively. 

\begin{table}[t]
\tbl{Summary of the theoretical results for the doublet
capture rate $\Gamma^D$ (in s$^{-1}$) of muon capture on 
deuteron. Only the calculations after the year 2000 are 
considered. The results obtained with the $^1S_0$ 
$nn$ final state are also shown. There where possible, the
theoretical uncertainty is also indicated.}
{\begin{tabular}{@{}ccc@{}} \toprule
Ref. & $\Gamma^D$ & $\Gamma^D(^1S_0)$ \\ \colrule
Ando {\it et al.}~\protect\cite{And02} & 386 & 254 $\pm$ 1 \\
Ricci {\it et al.}~\protect\cite{Ric10} & 423 $\pm$ 7 & 261 $\pm$ 7 \\
Marcucci {\it et al.}~\protect\cite{Mar11a} & 392 $\pm$ 2.3 & 248.6 $\pm$ 2.7 \\
\botrule
\end{tabular} \label{tab:res2}}
\end{table}

\begin{table}[b]
\tbl{Same as Table~\protect\ref{tab:res2} but for the
total capture rate $\Gamma_0$ of muon capture on 
$^3$He.}
{\begin{tabular}{@{}cc@{}} \toprule
Ref. & $\Gamma_0$ \\ \colrule
Marcucci {\it et al.}~\protect\cite{Mar02} & 1484 $\pm$ 8 \\
Gazit~\protect\cite{Gaz08} & 1499 $\pm$ 16 \\
Marcucci {\it et al.}~\protect\cite{Mar11a} & 1484 $\pm$ 13 \\
\botrule
\end{tabular} \label{tab:res3}}
\end{table}

Let us review our work of Ref.~\refcite{Mar11a}.
The first ingredient for any theoretical study of the 
reactions under consideration is the 
realistic Hamiltonian model used to describe the
initial and final $A=2$ and 3 
nuclear wave functions entering in Eqs.~(\ref{eq:h2ffz}) and~(\ref{eq:h3ffz}).
Two representative two-nucleon interaction models have been used, 
the phenomenological Argonne $v_{18}$ (AV18)~\cite{Wir95} 
and the potential derived within  
chiral effective field theory ($\chi$EFT) up to
next-to-next-to-next-to leading order (N3LO)
by Entem and Machleidt~\cite{Ent03}. 
These two models both reproduce the deuteron
observables and the large two-nucleon
scattering database with a $\chi^2$/datum $\simeq 1$.
Given the significant differences in their derivation and structure, 
they are believed to be 
a representative subset of the accurate two-nucleon interaction 
models available in the
literature. To accurately describe the $A=3$ nuclear systems,
it is well known that the two-nucleon potentials need to be augmented
by three-nucleon interactions. The Urbana IX (UIX)~\cite{Pud95} model
has been used in conjunction with the AV18, and the 
chiral three-nucleon interaction, derived up to
next-to-next-to leading order (N2LO) in Ref.~\refcite{Nav07}, has been used 
together with the N3LO.
The hyperspherical-harmonics (HH) method has been used to solve the $A$-body
bound and scattering problem, also
in the context of $A=2$ systems, for which of course
wave functions could have been obtained by direct solution
of the Schr\"odinger equation. 
The HH method for $A\geq 3$ 
has been reviewed in considerable detail in a series
of recent publications~\cite{Kie08,Viv06,Mar09}.

The weak current consists of polar- and axial-vector components, 
derived within two different frameworks, 
the ``Standard Nuclear Physics Approach'' (SNPA) and $\chi$EFT.
The first one goes beyond the impulse approximation, by including 
meson-exchange currents (MEC's) and terms arising
from the excitation of $\Delta$-isobar degrees of freedom.
The second approach includes two-body contributions
derived in 
heavy-baryon chiral perturbation theory 
(HB$\chi$PT) within a systematic expansion, up to 
N3LO~\cite{Par03,Son09}.
Since the transition operator matrix elements
are calculated using phenomenological wave functions, this second
approach is a ``hybrid'' $\chi$EFT approach ($\chi$EFT*). 
Here we briefly review the main characteristics of the weak current 
operator, both within SNPA and $\chi$EFT*. We consider
only the contributions beyond the one-body term, as the
one-body operators can be easily obtained performing a 
non-relativistic reduction of the single-nucleon weak current
of Eq.~(\ref{eq:wknuclcurr}), retaining corrections up to
order $(q^2/M_N^2)$~\cite{Mar11a,Mar00}.

The polar (scalar) weak current (charge) operator is related 
to the isovector part of the electromagnetic
current (charge) via the CVC hypothesis.
In SNPA, no free parameters are present in the model for 
the electromagnetic operator, which is able to 
reproduce the trinucleon magnetic moments to better than
1 \%~\cite{Mar11a}, as well as a large variety of
electromagnetic observables~\cite{Car98,Mar05,Mar08}.
In the case of $\chi$EFT*, no two-body contributions to the
scalar charge operator are present at N3LO, while
the vector current is decomposed into four terms~\cite{Son09}: 
the soft one-pion exchange ($1\pi$) term, vertex corrections to the one-pion
exchange ($1\pi C$), the two-pion exchange ($2\pi$), and a contact-term
contribution. Their explicit expressions can be found in 
Ref.~\refcite{Son09}. All the
$1\pi$, $1\pi C$ and $2\pi$ contributions contain low-energy constants
(LEC's) estimated using resonance saturation arguments, and
Yukawa functions obtained by performing the Fourier transform from 
momentum- to coordinate-space with a Gaussian regulator characterized
by a cutoff $\Lambda$. 
This cutoff determines the momentum scale below which
these $\chi$EFT currents are expected to be valid, i.e. 
$\Lambda$=500$\div$800 MeV~\cite{Par03}.
The contact-term electromagnetic contribution
is given as sum of two terms, isoscalar and isovector, each one 
with a LEC in front ($g_{4S}$ and $g_{4V}$), fixed to
reproduce the experimental values of $A=3$ magnetic 
moments. The resulting LEC's are given in Table V of Ref.~\refcite{Mar11a},
and listed again in Table~\ref{tab:lec} for completeness. 
The uncertainties on $g_{4S}$ and $g_{4V}$ are not due 
to the experimental errors on the triton and $^3$He 
magnetic moments, which are in fact negligible, rather to numerics. 

\begin{table}[b]
\tbl{The LEC's $g_{4S}$ and $g_{4V}$ associated with 
the isoscalar and isovector contact terms in the electromagnetic current, 
and the LEC $d_R$ 
of the two-body axial-current contact term, calculated for three
values of the cutoff $\Lambda$ with
triton and $^3$He wave functions obtained from the AV18/UIX model.
For $\Lambda=600$ MeV, 
the N3LO/N2LO model is also used.}
{\begin{tabular}{@{}ccccc@{}} \toprule
   & $\Lambda$ (MeV) & $g_{4S}$ & $g_{4V}$ & $d_R$ \\ \colrule
   & 500 & 0.69$\pm$0.01 & 2.065$\pm$0.006 & 0.97$\pm$0.07 \\
AV18/UIX 
   & 600 & 0.55$\pm$0.01 & 0.793$\pm$0.006 & 1.75$\pm$0.08 \\
   & 800 & 0.25$\pm$0.02 & --1.07$\pm$0.01 & 3.89$\pm$0.10 \\ \colrule
N3LO/N2LO
   & 600 & 0.11$\pm$0.01 & 3.124$\pm$0.006 & 1.00$\pm$0.09 \\
\botrule
\end{tabular}\label{tab:lec}}
\end{table}

The two-body axial current operators in SNPA as used in Ref.~\refcite{Mar11a},
as well as in the studies of the $pp$ and $hep$ 
reactions~\cite{Sch98,Mar00}, 
can be divided in two
classes: the operators of the first class
are derived from $\pi$- and $\rho$-meson exchanges and the 
$\rho\pi$-transition mechanism.  These mesonic operators
give rather small contributions~\cite{Mar11a}.
The operators in the second class are those that give
the largest two-body contributions, and are due to
$\Delta$-isobar excitation~\cite{Sch98,Mar00}. In particular,
in the dominant $N$-to-$\Delta$-transition axial current,
the $N$-to-$\Delta$ axial coupling constant ($g_A^*$) is retained as a 
parameter and 
is determined by fitting the experimental 
Gamow-Teller matrix element of tritium $\beta$-decay (GT$^{\rm EXP}$).
Also the pseudoscalar term in the $N$-to-$\Delta$-transition 
axial current is retained. 
It is important to note that the value of $g_A^*$ depends
on how the $\Delta$-isobar degrees of freedom are treated.
In the muon capture studies presented here, the two-body 
$\Delta$-excitation axial operator is
derived in the static $\Delta$ approximation, using first-order
perturbation theory. 
This approach
is considerably simpler than that adopted in Ref.~\refcite{Mar00},
where the $\Delta$ degrees of freedom were treated non-perturbatively,
within the so-called transition-correlation operator approach,
by retaining them explicitly in the nuclear wave functions~\cite{Sch92}.
The results for $g_A^*$ 
obtained within the two schemes differ by more 
than a factor of 2~\cite{Mar00}, but
the results for the observables
calculated consistently within the two different approaches are typically
within 1 \% of each other.
To be noticed that the presented 
SNPA two-nucleon weak current is not the only model
available in the literature. In fact, in Ref.~\refcite{Ric10}, 
two-body MEC's are derived from the hard pion chiral Lagrangians of the 
$N\Delta\pi\rho\omega a_1$ system, and are not constrained
to reproduce any experimental observable, like GT$^{\rm EXP}$. 
This is typically responsible for large model-dependence in the 
results, as some of the coupling constants and cutoff parameters entering the 
axial current are poorly known. 

The two-body axial current operator in $\chi$EFT 
consists of two contributions:
a one-pion exchange term and a two-nucleon 
contact-term. The explicit expressions for these terms 
can be found in Ref.~\refcite{Par03}.
While the coupling constants which appear in the one-pion exchange
term
are fixed by $\pi N$ data, the LEC which 
determines the strength of the contact-term ($d_R$) has been fixed
by reproducing GT$^{\rm EXP}$. 
The values of $d_R$ for 
$\Lambda$=500$\div$800 MeV are given in Table~\ref{tab:lec}~\cite{Mar11a}.
The experimental error on 
GT$^{\rm EXP}$ is primarily responsible for the
uncertainty in $d_R$.

Our results of Ref.~\refcite{Mar11a} for reaction~(\ref{eq:mud}) 
are compared in Table~\ref{tab:res2} with those of two 
previous calculations,
performed in SNPA~\cite{Ric10} and 
$\chi$EFT*~\cite{And02}. 
The first one uses the  Nijmegen I and 
Nijmegen 93~\cite{Sto94} Hamiltonian models to obtain
the nuclear wave functions, and 
MEC's derived from the Lagrangians of the 
$N\Delta\pi\rho\omega a_1$ system.
The second calculation 
uses the AV18~\cite{Wir95} potential to derive the wave functions,
and the same $\chi$EFT weak current model presented above, 
constrained to reproduce GT$^{\rm EXP}$ in tritium $\beta$-decay. 
However, the $1\pi C$, $2\pi$ 
and contact-term contributions to the weak vector current 
are not included. 
Furthermore, only the $S$-wave contribution in the $nn$
final scattering state (the $^1S_0$ state) is retained, 
and higher partial-wave 
contributions are estimated based on Ref.~\refcite{Tat90}. 
By inspection of Table~\ref{tab:res2} we can conclude
that: (i) our calculated $\Gamma^D$ values are in good agreement
with the results of Ando {\it et al.}~\cite{And02},
and the small existing differences has been traced back
to the inclusion in the weak vector current of the $1\pi C$, $2\pi$ 
and contact-term contributions~\cite{Mar11a}.
(ii) The calculated $\Gamma^D$ value of Ricci {\it et al.}~\cite{Ric10}
differs from the other results by 
7$\div$10 \%. 
In order to investigate the origin of the discrepancies between
our results~\cite{Mar11a} and those of Ando {\it et al.}~\cite{And02} 
on one side, and the results of Ricci {\it et al.}~\cite{Ric10} on the other,
we have repeated~\cite{Mar11b} the calculation of $\Gamma^D$ (and $\Gamma_0$), 
including the so-called 
``potential currents'', i.e. those operators arising 
when PCAC is implemented at the two-body level. 
It was argued in fact by Ricci {\it et al.}~\cite{Ric10} 
that ``omitting the potential current causes
an enhancement of the doublet transition rate $\Lambda_{1/2}$ [i.e.
$\Gamma^D$] by $\simeq$ 1 \%''. 
These currents were first constructed in Ref.~\refcite{Mos05},
and we have recently reviewed them in Ref.~\refcite{Mar11b}, 
where their explicit expression 
can be found. The calculation has been performed within $\chi$EFT*,
using the AV18 (AV18/UIX for $A=3$)
Hamiltonian model. Again the LEC which 
determines the strength of the axial current contact-term has been fixed
by reproducing GT$^{\rm EXP}$. The results for 
$\Gamma^D$ (and $\Gamma_0$) are 393.2 $\pm$ 0.8 s$^{-1}$ 
(1488 $\pm$ 9 s$^{-1}$), with $\Gamma^D(^1S_0)$=250.1 $\pm$ 0.8 s$^{-1}$, 
in perfect agreement with our previous results of 
Ref.~\refcite{Mar11a}. From this we can conclude that the 
potential currents proposed by Ricci {\it et al.}~\cite{Ric10} 
give negligible contributions to the 
rate $\Gamma^D$ (and $\Gamma_0$), and the 
discrepancy between the theoretical calculations is still a puzzling
problem.
From a historical point of view, it should be noticed that
such a discrepancy between different theoretical results
for $\Gamma^D$ already existed in the 
early nineties~\cite{Tat90,Ada90,Doi90-91}.
In fact, Adam and Truhl\`ik ~\cite{Ada90} found 
$\Gamma^D=$ 416 $\pm$ 7 s$^{-1}$, while Tatara {\it et al.}~\cite{Tat90} 
and Doi {\it et al.}~\cite{Doi90-91} found  
$\Gamma^D=$ 399 s$^{-1}$ and 402 s$^{-1}$, 
respectively. 
%Note that one of the authors of Ref.~\refcite{Ada90}
%is also author of Ref.~\refcite{Ric10}, and similarly one of the
%authors of Ref.~\refcite{Tat90} is also author of Ref.~\refcite{And02}.

Finally, we should also mention 
that a calculation of $\Gamma^D$ has been performed
within pionless EFT by Chen {\it et al.}~\cite{Che05}. 
The objective of this work,
however, is not to predict $\Gamma^D$, but rather to
find the relation between the two-nucleon axial current matrix element
entering the muon capture rate on deuteron and the $pp$ weak capture.
Within this approach, therefore, a precise experimental determination 
of $\Gamma^D$ will
put a stringent constraint on this matrix element, and consequently
on the $pp$ weak capture rate.

Using  
$A=3$ nuclear wave functions derived from the 
AV18/UIX or N3LO/N2LO Hamiltonian models, and the same SNPA or $\chi$EFT*
weak charge and current operators presented above, 
we have studied also 
the total capture rate $\Gamma_0$
for reaction~(\ref{eq:mu3})~\cite{Mar11a}.
The results are shown in Table~\ref{tab:res3}, and are compared
with other theoretical works of the last ten years~\cite{Mar02,Gaz08}.
Our calculation of Ref.~\refcite{Mar02} represents
the first attempt to study muon capture on $^3$He 
in a way that is consistent with 
the approach adopted for the weak proton capture reactions 
$pp$ and $hep$~\cite{Sch98,Mar00}.
The nuclear wave functions were obtained, 
within the HH method,  
from the AV18/UIX Hamiltonian model, and 
the nuclear weak current was derived within the SNPA, as presented above.
The theoretical uncertainty reported in 
Table~\ref{tab:res3} for $\Gamma_0$ 
results from the adopted 
fitting procedure and experimental error on GT$^{\rm EXP}$. 
Note that a calculation based on 
the older Argonne $v_{14}$ (AV14)~\cite{Wir84} two-nucleon and
Tucson-Melbourne (TM)~\cite{Coo79} Hamiltonian model yielded a $\Gamma_0$ of 
1486 $\pm$ 8 s$^{-1}$, suggesting a weak model-dependence.
In fact, we have demonstrated~\cite{Mar02}
that $\Gamma_0$ roughly scales as the triton binding energy. 
Therefore, any meaningful comparison between results obtained using different 
Hamiltonian models requires the inclusion of three-nucleon forces.
This is the reason why we have not considered in Table~\ref{tab:res3}
the results of Ho {\it et al.}~\cite{Ho02}, obtained, within the SNPA, 
without the inclusion of MEC's and, most important, three-nucleon 
interaction. 
In Ref.~\refcite{Mar02} we provide also the only available recent theoretical
prediction for the recoil asymmetry $A_v$, found to be 0.5350 $\pm$ 0.0014
with the AV18/UIX, in agreement with the experimental result of
Ref.~\refcite{Sou98}, 0.63 $\pm$ 0.09 (stat.)$^{+0.11}_{-0.14}$ (syst.). 
The results for $A_v$ are very little model-dependent,
but very sensitive to $G_{PS}(q^2)$: $A_v$ would vary by roughly 20 \%, if 
$G_{PS}(q^2)$ would be 50 \% larger than the PCAC value 
(see Fig.~1 of Ref.~\refcite{Mar02}). The corresponding 
variation for $\Gamma_0$ would be of the order of 5 \%.

The first study of reaction~(\ref{eq:mu3}) within $\chi$EFT* approach 
has been performed by Gazit~\cite{Gaz08}.
The nuclear wave functions have been obtained with the
Effective Interaction HH method~\cite{EIHH},
and the $\chi$EFT weak current presented above.
However, as in Ref.~\refcite{And02}, no $1\pi C$, $2\pi$ 
and contact-term contributions to the weak vector current 
are retained. 
The theoretical uncertainty reported in Table~\ref{tab:res3}
has two main sources: the experimental uncertainty on the triton half-life, 
and the calculation of electroweak radiative corrections~\cite{Cza07}.
Few comments are here in order: (i)
electroweak radiative corrections were not
included in our studies of Refs.~\refcite{Mar11a}
and~\refcite{Mar02} .
Were to be included, the central value
for $\Gamma_0$ would become 1493 s$^{-1}$ for both calculations, 
in nice agreement with
the result of Gazit~\cite{Gaz08}. (ii) The comparison between 
our study of 
Ref.~\refcite{Mar02} and that of Gazit~\cite{Gaz08} 
suggests that the SNPA and $\chi$EFT* results nicely
agree, when the MEC's are constrained to 
reproduce GT$^{\rm EXP}$. We have verified 
this observation\cite{Mar11a}
for both reaction~(\ref{eq:mud}) and~(\ref{eq:mu3}). However,
we have shown that 
$1\pi C$, $2\pi$, and contact terms in the
mesonic $\chi$EFT vector current are important
in order to achieve such an agreement.
If they were to be neglected,
$\Gamma_0$ would be 1453 s$^{-1}$~\cite{Mar11a}. 

Finally, we recall the studies of the early nineties
by Congleton and Fearing~\cite{Con92} and Congleton and Truhl\`ik~\cite{Con96}.
In the latter work, 
the nuclear wave functions were obtained from the AV14/TM
Hamiltonian model and the nuclear 
weak current retained contributions similar to those of Ref.~\refcite{Ric10}. 
The value obtained for the total capture rate $\Gamma_0$ 
was 1502 $\pm$ 32 s$^{-1}$, the uncertainty due to poor knowledge 
of coupling constants and cutoff parameters.

Only very recently, the first steps to study muon capture
reactions in a consistent $\chi$EFT framework have been
done~\cite{Mar11c,Ada11}. Although the results are not yet published,
the main ingredients of a $\chi$EFT calculation are outlined.
In particular, in Ref.~\refcite{Mar11c}, we have used the 
N3LO and N3LO/N2LO
interaction models and the $\chi$EFT weak current operator
presented above. Furthermore, the LEC $d_R$
determining the strength of the axial current contact-term, and 
the LEC $c_D$, entering the contact-term three-nucleon interaction
at N2LO, have been related, as suggested in Refs.~\refcite{Gar06}
and~\refcite{Gaz09}, as
\begin{equation}
d_R=\frac{M_N}{\Lambda_\chi g_A}c_D +\frac{1}{3}M_N(c_3+2c_4)+\frac{1}{6} \ ,
\label{eq:drcd}
\end{equation}
where
$c_3$ and $c_4$ are the LEC's of the
$\pi N$ Lagrangian, already part of the 
chiral two-nucleon potential at NLO,
and $\Lambda_\chi=700$ MeV is the 
the chiral-symmetry-breaking scale.
Then, the calculation is implemented in the following steps:
(i) all the LEC's present in the interaction
and in the current are set consistently, and the same 
cutoff regulator, i.e. ${\rm exp}[-(q/\Lambda)^4]$, is used both in the
current and in the N2LO three-nucleon interaction. 
(ii) The $^3$H and $^3$He ground state wave functions are calculated,
within the HH method, using the N3LO/N2LO Hamiltonian model, for two
values of $\Lambda=500$ and 600 MeV~\cite{Ent03,Mac11}, 
and the set of values $\{c_D,c_E\}$ are determined, 
for which the $A=3$ experimental binding energies (BE's) are reproduced.
A wide range of $c_D$ values has been spanned, and
in correspondence to each $c_D$ in this range, 
$c_E$ has been fixed in order to reproduce
either BE($^3$H) or BE($^3$He). 
(iii) For each set of $\{c_D,c_E\}$, the triton and
$^3$He wave functions are calculated and, using the
$\chi$EFT axial weak current discussed above, the Gamow-Teller
matrix element of tritium $\beta$-decay (GT$^{\rm TH}$) is
determined. This allows to determine the range of $c_D$ values for which
${\rm GT}^{\rm TH}={\rm GT}^{\rm EXP}$ within the experimental
error. A corresponding range for $c_E$ is given from
the previous step.
(iv) For the minimum and maximum values of $\{ c_D ; c_E \}$
in the selected range, the 
LEC's $g_{4S}$ and $g_{4V}$ entering the two-nucleon
contact terms of the electromagnetic current, and 
therefore the weak vector current, are determined 
by reproducing the $A=3$ magnetic moments.
At this point, the potential and current models are fully constrained,
and the 
results for $\Gamma^D$ and $\Gamma_0$ are $\chi$EFT predictions.
They are found to be $\Gamma^D=399 \pm 3$ s$^{-1}$ 
($\Gamma^D(^1S_0)=255 \pm 1$ s$^{-1}$) and 
$\Gamma_0=1494 \pm 21$ s$^{-1}$, including 
electroweak radiative corrections~\cite{Cza07}.
These results are in good 
agreement with the ones of the other calculations mentioned
above, except for
that of Ricci {\it et al.}~\cite{Ric10}, for which the 
discrepancy remains of the order of 4$\div$9 \%. 
On the other hand, in a similar calculation, Adam {\it et al.}~\cite{Ada11}
have found $\Gamma^D$
in the range 401.2 s$^{-1}$$\div$436.6 s$^{-1}$,
depending on the $\chi$EFT two-nucleon potential used.

We conclude remarking that 
a comparison between the calculated and measured 
rates for muon capture on $^3$He makes it possible to put a 
constraint on the induced pseudoscalar form factor $G_{PS}(q^2)$ at 
$q_0^2=-0.954\, m_\mu^2$, 
relevant for this reaction.  A similar comparison could be done for the
muon capture on deuteron. However, being the available experimental data so 
uncertain, such a comparison would be less significant.
Within the $\chi$EFT approach,
we have varied $G_{PS}(q_0^2)$ 
to match the theoretical upper (lower) value
with the experimental lower (upper) value for $\Gamma_0$~\cite{Mar11c}. 
This has allowed
to obtain for $G_{PS}(q_0^2)= 8.2 \pm 0.7$, 
in very good agreement with the $\chi$PT prediction 
of Eq.~(\ref{eq:gpsth}), which gives
$G_{PS}^{\rm TH}(q_0^2)=7.99 \pm 0.20$~\cite{GPth1}.

\section{Conclusions}
\label{sec:concl}

Muon capture reactions on light nuclei, in particular deuteron
and $^3$He, have demonstrated to be an interesting, fruitful
and controversial field of research, both experimentally and
theoretically. The work on this subject has been extensive, 
and the last few years have seen 
even a growth of interest and research. At this point, the experimental
situation can be summarized as follows: (i) the total 
rate for muon capture on $^3$He, reaction~(\ref{eq:mu3}), is very
well determined, with an accuracy of 0.3 \%, hard to be reached
by any present theoretical calculation. (ii) The angular
correlation parameters, or so-called recoil asymmetries, are 
poorly known. Only the vector asymmetry $A_v$ has been measured,
but the experimental error is still very large, of the 
order of $\sim 30$ \%. (iii) The rate
for muon capture on deuteron, reaction~(\ref{eq:mud}), from 
the doublet hyperfine state is also poorly known, with experimental
values which agree among each other, but have uncertainties of
6$\div$10 \%. However, 
the ongoing experiment performed by the MuSun 
Collaboration at 
PSI~\cite{And10,Kam10} will clarify the situation
and determine the rate with a precision of 1.5 \%.

The theoretical situation is evolving very fast, and a large 
effort has been put in the past few years to reduce
as much as possible the theoretical uncertainty on
the calculated observables. In particular, it has been 
shown~\cite{Mar11a,Mar02} that
the model-dependence, 
relative to the adopted models for the nuclear interaction and 
weak currents, can be strongly reduced by fitting the unknown
parameters of the nuclear currents to some significant observables,
as the tritium half-life and the $A=3$ magnetic moments. 
A crucial role in these calculations
is played by the numerical techniques used to calculate the
few-body wave functions, with the considered accurate (and highly complex)
Hamiltonian models. Without such a fundamental ingredient, all the
calculations mentioned above would be affected by
a much larger uncertainty. 
However, significant discrepancies remain between the available
theoretical calculations of the rate for muon capture on deuteron,
reaction~(\ref{eq:mud}), and their origin is still 
to be understood.
Finally, the first steps toward a $\chi$EFT, and ultimately
QCD-based, prediction have been made~\cite{Mar11c,Ada11}. 

Considering the two main motivations to study muon capture on light nuclei,
i.e. (i) to provide significant tests for the theoretical frameworks
used in the study of reactions of astrophysical interest not accessible 
experimentally, and (ii) to extract the value for the 
induced pseudoscalar form factor $G_{PS}(q^2)$ and ultimately validate
the $\chi$PT predictions, further investigations are highly recommended.
In particular, there are very few studies on the muon capture 
reactions on $^3$He in the two- and three-body breakup channels,
both theoretically~\cite{Ski99} and experimentally~\cite{Kuh94,Bys04}.
Furthermore, an accurate
measurement for the angular
correlation parameters of muon capture on $^3$He
could put even a more stringent constraint on $G_{PS}(q^2)$.

\section*{Acknowledgments}

I would like to thank my collaborators
L.\ Girlanda, A.\ Kievsky, M.\ Piarulli, S.\ Rosati, R.\ Schiavilla,
and M.\ Viviani, for their input and many contributions to the 
subject reviewed here. I also would like to thank P.\ Kammel for
useful discussions, especially on the experimental aspects of muon capture
reactions.

\end{document}